\documentclass[prd,nofootinbib,preprintnumbers]{revtex4}
\usepackage{latexsym}
\usepackage{dcolumn}
\usepackage{epsfig} 
\def\beq{\begin{equation}}
\def\eeq{\end{equation}}
\def\bea{\begin{eqnarray}}
\def\eea{\end{eqnarray}}
\def\bq{\begin{quote}}
\def\eq{\end{quote}}
\def\ben{\begin{enumerate}}
\def\een{\end{enumerate}}
\def\nn{\nonumber}
\def\fr{\frac}

\def\ra{\rightarrow}

\def\dm{\Delta m}
\def\dtm{\Delta \tilde{m}}

\def\lesssim{\mathrel{\mathpalette\vereq<}}
\def\gtrsim{\mathrel{\mathpalette\vereq>}}
\makeatletter
\def\vereq#1#2{
\lower3pt\vbox{\baselineskip1.5pt \lineskip1.5pt
\ialign{$\m@th#1\hfill##\hfil$\crcr#2\crcr\sim\crcr}}}
\makeatother

\def\beq{\begin{equation}}
\def\eeq{\end{equation}}
\def\bea{\begin{eqnarray}}
\def\eea{\end{eqnarray}}
\def\bq{\begin{quote}}
\def\eq{\end{quote}}
\def\ben{\begin{enumerate}}
\def\een{\end{enumerate}}
\def\nn{\nonumber}
\def\fr{\frac}

\def\ra{\rightarrow}

\begin{document}

\preprint{FERMILAB-Pub-02/217-T}

\title{Long Baseline Neutrino Experiments and the LOW solution -- 
What is Left to Do and How Well Can It Be Done}

\author{Gabriela Barenboim}
\affiliation{Theoretical Physics Division, Fermilab, 
P.O. Box 500, Batavia, IL, 60510-0500, USA }
\author{Andr\'e de Gouv\^ea}
\affiliation{Theoretical Physics Division, Fermilab, 
P.O. Box 500, Batavia, IL, 60510-0500, USA }

\begin{abstract}
The neutrino community is convinced that 
if the LMA solution is indeed correct, the prospects for next-generation,
long-baseline neutrino experiments are very exciting. 
In this work we will argue that the prospects are not less exiting
if the answer to the solar neutrino puzzle lies in the LOW region.
For this purpose we explore the oscillation probabilities which are accessible 
to experiments which employ conventional neutrino beams. 
We consider the electron (anti)neutrino 
appearance channel, which provides information regarding not only the small 
$U_{e3}$-element of the leptonic mixing matrix but also the neutrino 
mass hierarchy, and also include information, which should come from  
the muon (anti)neutrino disappearance channel, 
regarding the magnitude of the atmospheric mass-squared difference
($|\Delta m^2_{13}|$) and mixing angle ($\theta_{23}$). 
We comment on the presence of ambiguities and the experiments needed 
to solve them.
\end{abstract}
%

\maketitle

\section{Introduction}

There is unambiguous evidence of new physics in the leptonic
sector of the standard model. Furthermore, all experimental neutrino data 
\cite{solar,SNO_solar,atm,atm_sk,lsnd}
are consistent with the hypothesis that neutrinos have mass and mix. This being
the case, the goal of this and the next generation of neutrino experiments is to
determine the neutrino masses and the leptonic mixing matrix to the best of
their capabilities. 

%

One of the issues which is to be resolved by the present generation
of neutrino experiments involves the determination of the so-called
solar parameters, $\Delta m^2_{12}$ and $\tan^2\theta_{12}$. The current data 
strongly require $\tan^2\theta_{12}$ to be of order unity, while 
$10^{-4}~{\rm eV^2}\lesssim\Delta m^2_{12}\lesssim 10^{-5}$~eV$^2$ (the large mixing
angle solution (LMA)) or $\Delta m^2_{12}\ll 10^{-6}$~eV$^2$ 
\cite{SNO_solar,solar_fits,general_fit}. 
The latter will be referred
to in this paper as the LOW solution.\footnote{The current data allow, at the ``three sigma''
level, not only the LMA solution, but also the LOW, QVO, and VAC
solutions, depending on the data analysis. 
See, \cite{SNO_solar,solar_fits,general_fit} for details.} While global analyses 
indicate that
the LMA solution is significantly preferred, it is fair to
say that the LOW solution still remains a realistic possibility \cite{fit_strumia}. 

In any case, the KamLAND reactor neutrino experiment \cite{Kamland}  will, in the near future,
definitively establish whether the LMA solution is driving solar 
neutrino oscillations. If KamLAND observes an oscillation signal, it will not only
exclusively select the LMA solution but also ``pin-point'' the value of 
solar parameters \cite{Kam_sim,Kam_sim_ex}. 
If KamLAND does not observe an oscillation signal, the LMA solution 
will be unambiguously (modulo CPT violation \cite {nos}) ruled out \cite{Kam_sim_ex}, and we will
have to wait for the Borexino solar neutrino experiment \cite{Borexino}, which is scheduled
to start data-taking in 2003, in order to definitively piece the solar neutrino puzzle 
\cite{borex_sim}.  

%

If the LMA solution is indeed correct, the prospects for next-generation,
long-baseline neutrino experiments are very exciting. Not only can these
experiments probe the remaining mixing angle of the leptonic mixing matrix,
but there remains the possibility of also observing CP-violating effects. This
scenario has been thoroughly studied, for different long-baseline experiments 
(superbeams, neutrino factories, etc), in the literature \cite{lit1,our_paper,JHF}.   

In this work we will explore the physics capabilities of future
long-baseline oscillation experiments if the LMA solution is not correct. 
We will argue that a rich physics program lies ahead despite
the fact that the door to leptonic CP violation will be closed (indeed, the study of any
``solar effect'' in terrestrial oscillation experiments becomes virtually impossible). 
First, we discuss the oscillation probabilities which are accessible 
to experiments which employ conventional neutrino beams (which are mostly 
composed of muon-type (anti)neutrinos). We consider both the electron (anti)neutrino 
appearance channel, which provides information regarding not only the small 
$U_{e3}$-element of the leptonic mixing matrix but also the neutrino mass hierarchy, 
and the muon (anti)neutrino disappearance channel, which provides information 
regarding the magnitude of the atmospheric mass-squared difference
($|\Delta m^2_{13}|$) and mixing angle ($\theta_{23}$). A comparison of both expressions
reveals that the information provided by the disappearance channel leads to ``ambiguities''
\cite{amb}
in the determination of the currently unknown oscillation parameters. Far from being a
nuisance, these degeneracies are an indication that more physical quantities can be
measured given the correct amount of information.  

Next, we simulate and analyse ``data'' from a finely segmented iron-scintillator detector located
off-axis from the future NuMI beam at Fermilab. In measuring the oscillation parameters,
we include the realistic uncertainties on all relevant parameters, and consider 
several scenarios in order to fully study all degeneracy issues. We find that one should
be able to determine the neutrino mass-hierarchy by combining information from neutrino
and antineutrino oscillations if $|U_{e3}|^2$ is large enough (solving one ``degeneracy 
issue''), but not whether $\theta_{23}$ is less than or greater than $\pi/4$. We comment
on how this degeneracy might be attacked. Finally, we analyse the issue of how well the 
atmospheric mass-squared difference should be known before
choosing the location of the off-axis detector.  

 

\section{The Disappearance Channel}
\label{sec_dis}

We are interested in baselines around $10^3$ km and
order 1~GeV neutrinos energies, chosen in order to explore the first 
oscillation peak. We hope to show, contrary to some claims, that
there is no need to go beyond the first peak in order to disentangle
the oscillation features. For this class of experiments, the parameter
$\alpha$ (see Appendix~A), which measures the strength of matter effects, is given by
\bea
\alpha\equiv \frac{2 \sqrt{2} G_F n_e E}
{\Delta m_{13}^2} \simeq \frac{2.8 \,\, 10^{-4}}{\Delta m_{13}^2}
\left( \frac{E}{\mbox{GeV}}\right),
\eea
where $n_e$ is the electron number density in the Earth and E is the energy
of the neutrino (see Appendix~A for definitions and conventions).

In the region of interest, $\alpha$ is relatively small ($\alpha \simeq .1-.2$)
such that,
up to quartic terms in $|U_{e3}|\equiv s_{13}$, 
the muon neutrino survival probability can be written as 
(complete formulas are given in Appendix~A) 
\bea
P\left(\nu_\mu \rightarrow \nu_\mu \right) &=& 1 - 4 s_{23}^2 c_{23}^2
c_{13}^4 
\sin^2 \left[ \Delta_{13} \right]- s_{23}^2 \frac{4 s_{13}^2}
{(1 - \alpha)^2}
 \sin^2\left[ \Delta_{13} (1 - \alpha )\right].
\label{mm}
\eea
Note that we have also expanded $\tilde{\Delta}_{13}=\Delta_{13}
|1-\alpha|(1+O(\alpha |U_{e3}|^2))$. 
Matter effects, therefore, only modify the vacuum survival probability by terms which
are $O(|U_{e3}|^2\alpha)$, and can safely be neglected. The vacuum expression can be
written as
\begin{eqnarray}
&P(\nu_\mu\rightarrow\nu_\mu)=
P(\bar{\nu}_\mu\rightarrow\bar{\nu}_\mu)
=1-4|U_{\mu3}|^2(1-|U_{\mu3}|^2)\sin^2(\Delta_{13}); \\
&|U_{\mu_3}|=\sin\theta_{23}\cos\theta_{13},
\end{eqnarray} 
meaning that one can measure $|\Delta m^2_{13}|$ (from the oscillation frequency) 
and $\sin^22\theta^{\rm eff}_{23}\equiv
4|U_{\mu3}|^2(1-|U_{\mu3}|^2)$ (from the magnitude of the oscillation). For $|U_{e3}|\ll 1$ 
\begin{equation}
\sin^22\theta_{23}^{\rm eff}=\sin^22\theta_{23}\left[1-|U_{e3}|^2\frac{2(1-\cos2\theta_{23})
\cos2\theta_{23}}{\sin^22\theta_{23}}+O(|U_{e3}|^4)\right].
\end{equation}
In the limit of vanishing $|U_{e3}|$, $\sin^22\theta_{23}^{\rm eff}=\sin^22\theta_{23}$, while
a precise enough extraction of $\sin^22\theta_{23}$ from
$\sin^22\theta_{23}^{\rm eff}$ is sensitive to the value of $|U_{e3}|$. 
Indeed, the fact that $|U_{e3}|^2$ is not known induces an irreducible
uncertainty on extracting $\sin^22\theta_{23}$ from $\nu_{\mu}$ disappearance experiments
as large as $\Delta(\sin^22\theta_{23})\sim 0.02$, depending on the value of 
$\sin^22\theta_{23}$ and the uncertainty on 
$|U_{e3}|^2$.\footnote{It is curious to note that for ``maximal mixing''
in the atmospheric sector ($\theta_{23}=\pi/4$), there are only $|U_{e3}|^4$ corrections.} 

The information needed in order to interpret the 
$\nu_e$ appearance channel (as will be discussed in the next section) 
is $\sin^2\theta_{23}$ (not $\sin^22\theta_{23}$) and 
($\Delta m^2_{13}$ (not $|\Delta m^2_{13}|$). This implies that, as far as the appearance channel
is concerned, the information from the disappearance channel is {\sl four-fold degenerate},
meaning that a unique measurement of $(\sin^22\theta^{\rm eff}_{23},
|\Delta m^2_{13}|)$ ``splits'' into 
\begin{equation}
(\sin^2\theta_{23},\Delta m^2_{13})=
\left(\frac{1}{2}\left(1\mp\sqrt{1-\sin^22\theta_{23}^{\rm eff}}\right),
\pm|\Delta m^2_{13}|\right),
\end{equation} 
where the subleading $|U_{e3}|^2$ effects have been ignored.

\begin{figure}[ht]
\vspace{1.0cm}
\centerline{\epsfxsize 14.2cm \epsffile{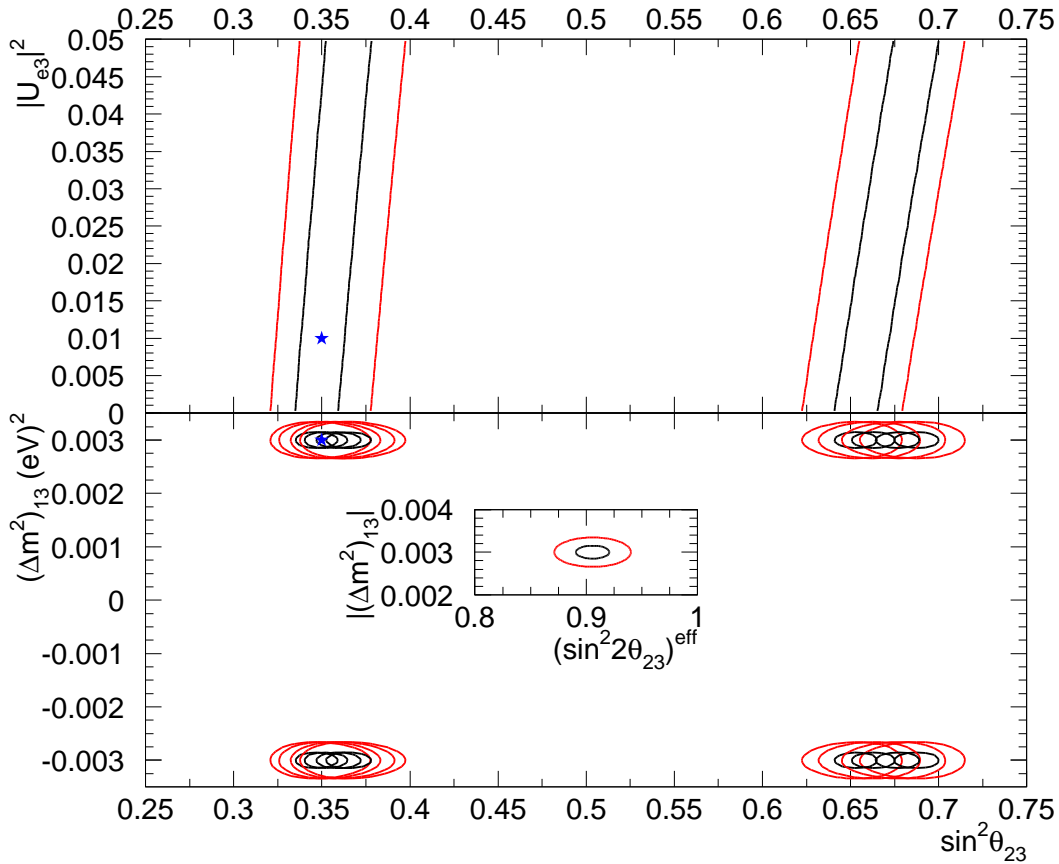}}
\caption{One and three sigma confidence level contours 
in the $(\sin^2\theta_{23}\times|U_{e3}|^2)$ and $(\sin^2\theta_{23}\times\Delta 
m^2_{13})$-planes, obtained from the ``measurement'' $|\Delta m^2_{13}|=(3\pm0.1)\times 
10^{-3}$~eV$^2$ and $\sin^22\theta_{23}^{\rm eff}=0.91\pm 0.01$, depicted in the inset located
in the center. See text for details.}
\label{dis_mea}
\end{figure}

This parameter degeneracy is depicted in Fig.~\ref{dis_mea}, where
we present the one and three sigma confidence level (CL) contours in the $(\sin^2\theta_{23}\times
|U_{e3}|^2)$ and $(\sin^2\theta_{23}\times\Delta m^2_{13})$-planes, assuming that one has measured
$|\Delta m^2_{13}|=(3\pm 0.1)\times 10^{-3}$~eV$^2$ and $\sin^22\theta^{\rm eff}_{23}=
0.91\pm0.01$ (to be exact, the input point corresponds to 
$\sin^2\theta_{23}=0.35$ and $|U_{e3}|^2=0.01$, as indicated by the (blue) stars in 
Fig.~\ref{dis_mea}). We also impose the CHOOZ bound \cite{chooz}
as a hard cutoff, $|U_{e3}|^2<0.05$, and
that the measurements of the effective atmospheric angle and the absolute value
of the atmospheric mass-squared difference are uncorrelated. This is a simplifying assumption,
which would be a good approximation if future long-baseline $\nu_{\mu}$ disappearance
experiments run at distances and energies which are around the minimum of the $\nu_{\mu}$
survival probability. In this case, and assuming a good energy resolution and calibration,
the mass-squared difference can be obtained by the position of the minimum, while the 
$\sin^22\theta_{23}^{\rm eff}$ is obtained by the magnitude of the suppression. 
The capabilities of long-baseline experiments to measure the atmospheric parameters is currently
under investigation \cite{investigation}, and clearly depend on the detector-type and 
the neutrino beam.

In each plot of Fig.~\ref{dis_mea}, the CLs are defined for two degrees of 
freedom, {\it i.e.,}\/ the absent parameter in each figure has been ``integrated out.'' 
On the bottom plot, we present the confidence ellipses for different fixed values of 
$|U_{e3}|^2$. The appropriate one and three sigma level regions are to be interpreted as 
the envelope of the different inner and outer ellipses, respectively. 

Several comments are in order. First, the alluded to four-fold degeneracy is clearly
visible in the bottom plot of Fig.~\ref{dis_mea}. Second, the fact that $|U_{e3}|^2$ is not 
known implies a significant increase of $\Delta(\sin^2\theta_{23})$, as can be observed 
by comparing each individual ellipse in the bottom plot with the size of the envelope. The
situation will improve once more information (from the appearance channel) 
regarding the magnitude of $U_{e3}$ is included.

To conclude this section, we point out the simple but important fact that a Gaussian measurement
of $\sin^22\theta_{23}^{\rm eff}$ does {\sl not} translate into a Gaussian measurement of
$\sin^2\theta_{23}$. This is true even when $|U_{e3}|=0$. This can be easily understood in the
following way. If the measurement of $\sin^22\theta$ is Gaussian, it can be transformed into a
$\chi^2$ function $(\sin^22\theta-x)^2/(\sigma_x)^2$, where $x=\sin^22\theta^*$ 
is the measured value of $\sin^2\theta$ with error $\pm\sigma_x$. This translates into
\begin{equation}
\chi^2(\sin^2\theta)=\frac{((\sin^2\theta-y_1)^2(\sin^2\theta-y_2)^2}{(\sigma_x/4)^2},
\end{equation} 
where $y_{1,2}$ are the solutions to $y(1-y)-x/4=0$. If $y_{1,2}$ are separated enough (in units
of $\sigma_x$) the $\chi^2$ splits into the sum of two parabola, with error $\sigma_y=
\sigma_x/|4(y_1-y_2)|$ (of course, $|y_1-y_2|=|\cos2\theta^*|$). This will not be the case of
maximal mixing ($y_1=y_2=0.5$), where the $\chi^2$ function will be a quartic 
monomial which cannot be approximated by a parabola. This implies that a measurement of 
$\sin^22\theta=1$ with an uncertainty of 0.01
will imply $\sin^2\theta=0.5\pm 0.05$, but the $\pm0.05$
one sigma uncertainty is very non-Gaussian, meaning that, for example, the three 
sigma error bar is not $\pm0.15$ (as a matter of fact, the three sigma error bar is significantly 
smaller than $\pm 0.15$). 

\section{The Appearance Channel}
\label{sec_app}

The electron neutrino appearance probability is given by (again, for small $\alpha$)
\bea
P\left(\nu_\mu \rightarrow \nu_e \right) &=& s_{23}^2 \frac{4 s_{13}^2}
{(1 - \alpha)^2}
 \sin^2\left[ \Delta_{13} (1 - \alpha )\right], 
\label{pmue}
\eea
while the approximate expression for antineutrinos is given by Eq.~(\ref{pmue}) if
one exchanges $\alpha\rightarrow-\alpha$. (For exact expressions see Appendix~A).
Note that Eq.~(\ref{pmue}) can be significantly
modified by the presence of matter (at $O(\alpha)$). 

The dominant effect will arise from the $(1-\alpha)^{-2}$ modification of the 
overall magnitude of the oscillation probability, since we assume that the available 
experimental information will not be able to properly ``see'' the position of the 
oscillation maximum.  This implies that the $\nu_{e}$ appearance rate {\sl for a fixed value 
of $|U_{e3}|^2$} is enhanced [suppressed] with respect to the pure vacuum one
if the neutrino mass hierarchy is normal ($\alpha>0$) [inverted, ($\alpha<0$)]. 
The opposite happens in the antineutrino channel. If the mass hierarchy is not known,
a degeneracy in the determination of $|U_{e3}|^2$ surfaces, as a fixed value of $|U_{e3}|^2$
and a normal mass hierarchy will yield the same number of $\nu_e$ appearance events within
some energy window as a larger value of $|U_{e3}|^2$ and an inverted mass hierarchy.
This ambiguity can be solved by, for example, 
combining the information obtained with the neutrino and
antineutrino channels (whose matter effects are opposite) or by changing
the baseline of the experiment, such that the significance of the matter effect is modified. 
Note that precise information regarding the location ({\it i.e.},\/ the incoming neutrino energy) 
of the minimum of the $\nu_{\mu}$ survival probability and the maximum of the 
$\nu_e$ appearance 
probability will also identify the neutrino mass hierarchy. This information, however, is hard
to obtain in the appearance channel.  

Another degeneracy ensues if $\sin^2\theta_{23}\neq 0.5$. This comes from the fact that
Eq.~(\ref{pmue}) is directly proportional to $\sin^2\theta_{23}$, while, as discussed
in Sec.~\ref{sec_dis} (see Fig.~\ref{dis_mea}) the disappearance channel cannot distinguish
$\sin^2\theta_{23}$ from $1-\sin^2\theta_{23}$ (ignoring small $|U_{e3}|^2$ corrections). 
This means that for every $(|U_{e3}|^2,\sin^2\theta_{23})$ pair there is another choice
for the oscillation parameters (namely, $(|U_{e3}|^2\tan^2\theta_{23},\cos^2\theta_{23})$) which
yields exactly the same $P(\nu_{\mu}\rightarrow\nu_e)$ as a function of energy, as long
as the approximations that go into writing Eq.~(\ref{pmue}) apply. 
Contrary to the ``hierarchy degeneracy,'' the degeneracy in the atmospheric angle 
cannot be solved by comparing conjugated channels, as it is not 
modified by the presence of matter. Similar experiments at different baselines and neutrino
energies will also have no effect. In order to eliminate this degeneracy it is necessary
to either look at different oscillation modes (which come with distinct 
$\theta_{23}$ dependences) or to probe $P(\nu_{\mu}\rightarrow\nu_e)$ around 
$\alpha=1$, where the value of $|U_{e3}|^2$ affects the $\nu_e$ appearance rate in 
a less trivial way. 

Candidates for different oscillation modes include
the search for the disappearance of electron-type neutrinos or $\nu_{e}\leftrightarrow\nu_{\tau}$
oscillations. For $|U_{e3}|^2\ll 1$
\bea
P\left(\nu_e \rightarrow \nu_e \right) &=& 1 - \frac{4 s_{13}^2}
{(1 - \alpha)^2}
 \sin^2\left[ \Delta_{23} (1 - \alpha )\right], \\
P\left(\nu_e \rightarrow \nu_\tau \right) &=& c_{23}^2 \frac{4 s_{13}^2}
{(1 - \alpha)^2}
 \sin^2\left[ \Delta_{23} (1 - \alpha )\right], 
\eea
which depend only on $|U_{e3}|^2$ or on $\cos^2\theta_{23}|U_{e3}|^2$. We will comment more on
this issue in the next section.

\section{Simulations of the Off-Axis Experiment(s)}
\label{sec_sim}

We simulate and analyse 
``data'' for electron (anti)neutrino appearance in a highly segmented iron-scintillator
detector located off-axis from the NuMI neutrino beam. The detector is described
in \cite{our_paper}, together with the off-axis neutrino beams and the reconstruction
efficiencies, and we refer readers to it for more details. We will always assume that the
data consists of 120 kton-years of running with a  ``neutrino beam'' (see 
\cite{our_paper}) plus, whenever applicable, 300 kton-years of running with an 
``antineutrino beam.'' We assume that the detector is located 12~km off-axis and 900~km away from 
the neutrino source (see \cite{our_paper}), unless otherwise noted.
We further include in the ``data'' analysis the fact that $|\Delta m^2_{13}|$ is measured
with a precision of $\Delta(|\Delta m^2_{13}|)=0.1\times 10^{-3}$~eV$^2$, while 
$\sin^22\theta_{23}^{\rm eff}$ is measured with a precision 
$\Delta(\sin^22\theta_{23}^{\rm eff})=0.01$. This type of  
precision has been quoted, for example, in studies of the physics capabilities of a future
JHF to SuperKamiokande neutrino program \cite{JHF}, 
and we assume that the study of muon disappearance at the
off-axis detector being considered here should yield similar precision. 
Both errors are considered to be one sigma, uncorrelated, Gaussian 
errors.\footnote{As noted in Sec.~\ref{sec_dis}, a Gaussian error
in $\sin^22\theta_{23}$ does not leads to a Gaussian error for $\sin^2\theta_{23}$. This
is appropriately taken into account.} 

In order to discuss all relevant physics issues, we will consider several scenarios:
$\sin^2\theta_{23}=0.35,0.5,0.65$, $\Delta m^2_{13}=\pm (2,3,4)\times10^{-3}$~eV$^2$ and
$|U_{e3}|^2<0.05$. These choices are meant to be close to the current best fit
of the atmospheric neutrino data \cite{atm_sk,general_fit} or at the boundaries of the
currently allowed 99\%~CL region.  
For concreteness, we have fixed $\Delta m^2_{12}=0.93\times 
10^{-7}$~eV$^2$, $\sin^2\theta_{12}=0.33$. It should be clear that any other LOW value
for the solar parameters would have yielded identical results.

\begin{figure}[ht]
\vspace{1.0cm}
\centerline{\epsfxsize 14.2cm \epsffile{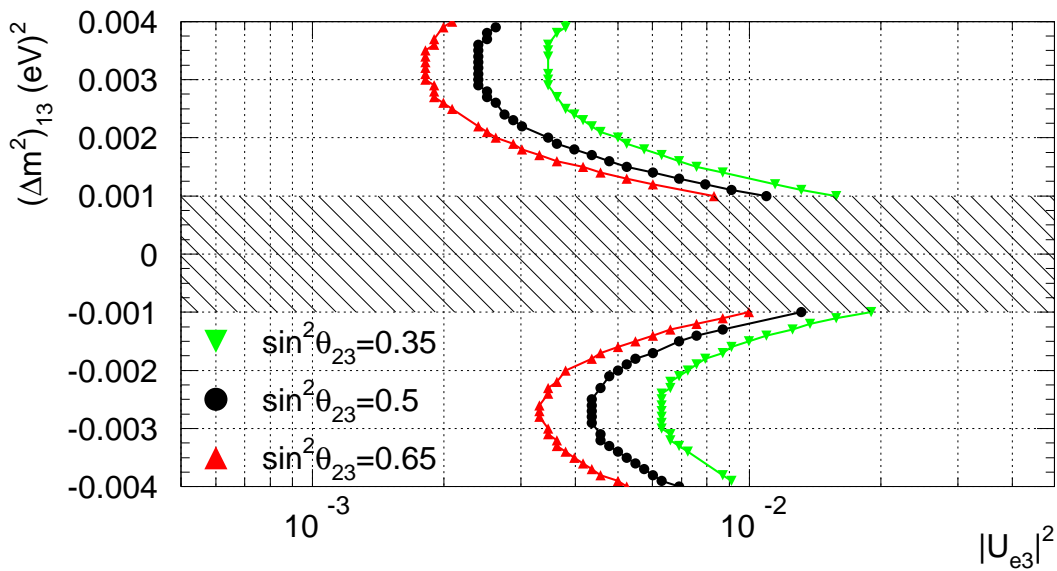}}
\caption{Three sigma confidence level sensitivity of 120 kton-years of ``neutrino beam'' 
data, 12~km off-axis and 90~km away from the NuMI beam, as a function of
$\Delta m^2_{13}$ and $|U_{e3}|^2$, for $\sin^2\theta_{23}=0.35,0.5,0.65$.
The hatched region is currently ruled out by the atmospheric neutrino data at the three sigma
confidence level.}
\label{sensitivity}
\end{figure}

First, we determine the capability of the setup in question to observe a signal. We do 
that by computing the sensitivity to $\nu_e$ appearance as a function of the relevant
oscillation parameters. The three sigma sensitivity CL curve for 120~kton-years
of neutrino-beam running (see \cite{our_paper} for details) is depicted in 
Fig.~\ref{sensitivity} in the $(|U_{e3}|^2\times\Delta m^2_{13})$-plane, for both neutrino
mass hierarchies (signs of $\Delta m^2_{13}$) 
and three different values of $\sin^2\theta_{23}$. By sensitivity we mean
that, if $|U_{e3}|^2,\Delta m^2_{13}$ correspond a point in the curve, the probability
that the number of events observed after 120 kton-years of running is due to background
is 0.27\%. 

As discussed in Sec.~\ref{sec_app}, the sensitivity is greater for larger values of 
$\sin^2\theta_{23}$ and for the normal neutrino mass hierarchy, due to a larger 
number of $\nu_e$-appearance candidates. Not surprisingly, Fig.~\ref{sensitivity} indicates
that the optimal sensitivity occurs for $\Delta m^2_{13}\simeq 3\times 10^{-3}$~eV$^2$, as we
have chosen the neutrino-beam energy profile ({\it i.e.,}\/ the off-axis distance) 
and experimental baseline in order to obtain
$\Delta_{13}\simeq\pi/2$ for this particular value of the mass-squared difference.  

\begin{figure}[ht]
\vspace{1.0cm}
\centerline{\epsfxsize 14.2cm \epsffile{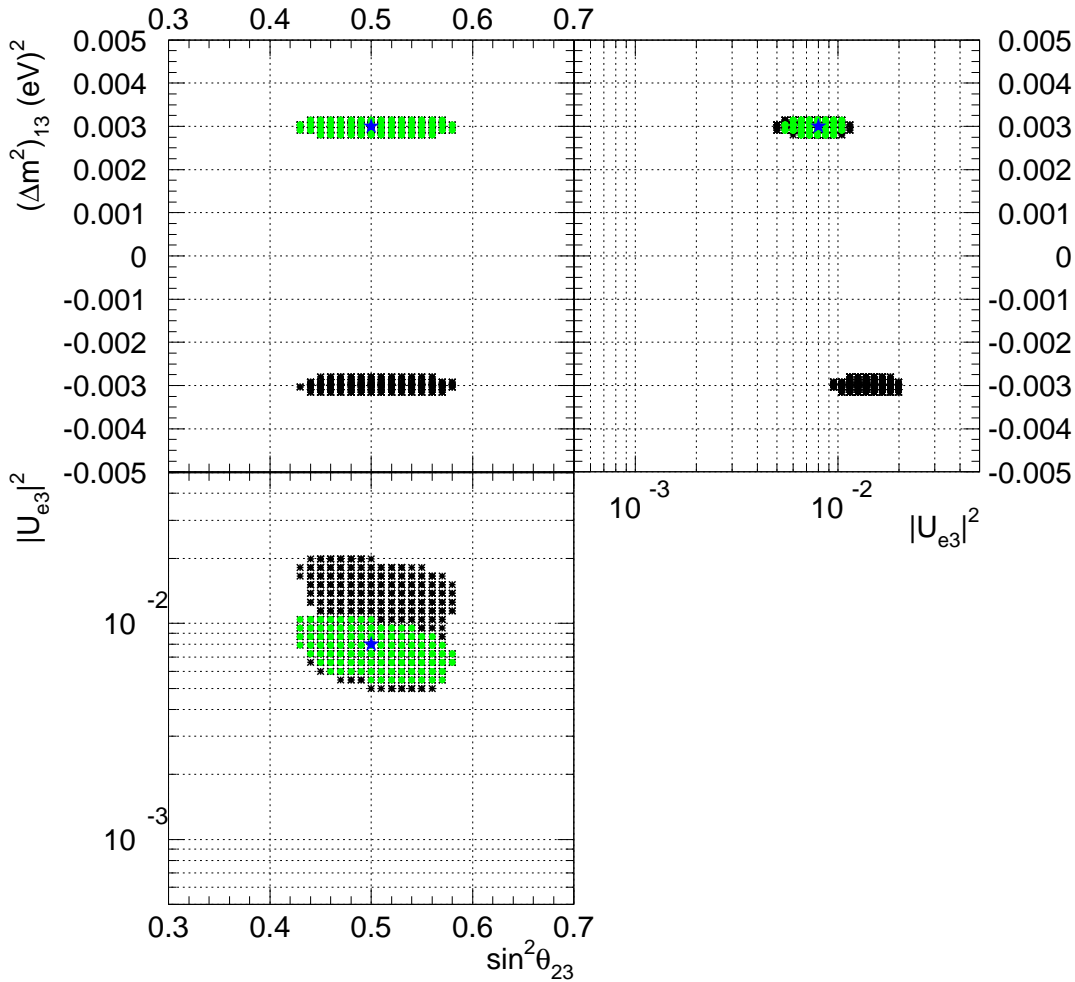}}
\caption{Two-sigma confidence level regions in the  $(\Delta m^2_{13}\times|U_{e3}|^2)$, 
$(|U_{e3}|^2\times\sin^2\theta_{23})$, and $(\Delta m^2_{13}\times\sin^2\theta_{23})$-planes
obtained after 120 kton-years of ``neutrino-beam data'' (dark squares) combined
with 300 kton-years of ``antineutrino-beam data'' (light [green] circles) for the
following input value for the mixing parameters (indicated by the [blue] stars): 
$\Delta m^2_{13}=3\times 10^{-3}$~eV$^2$, $|U_{e3}|^2=0.008$, and $\sin^2\theta_{23}=0.5$. 
See text for details. }
\label{mea_05}
\end{figure}

\begin{figure}[ht]
\vspace{1.0cm}
\centerline{\epsfxsize 14.2cm \epsffile{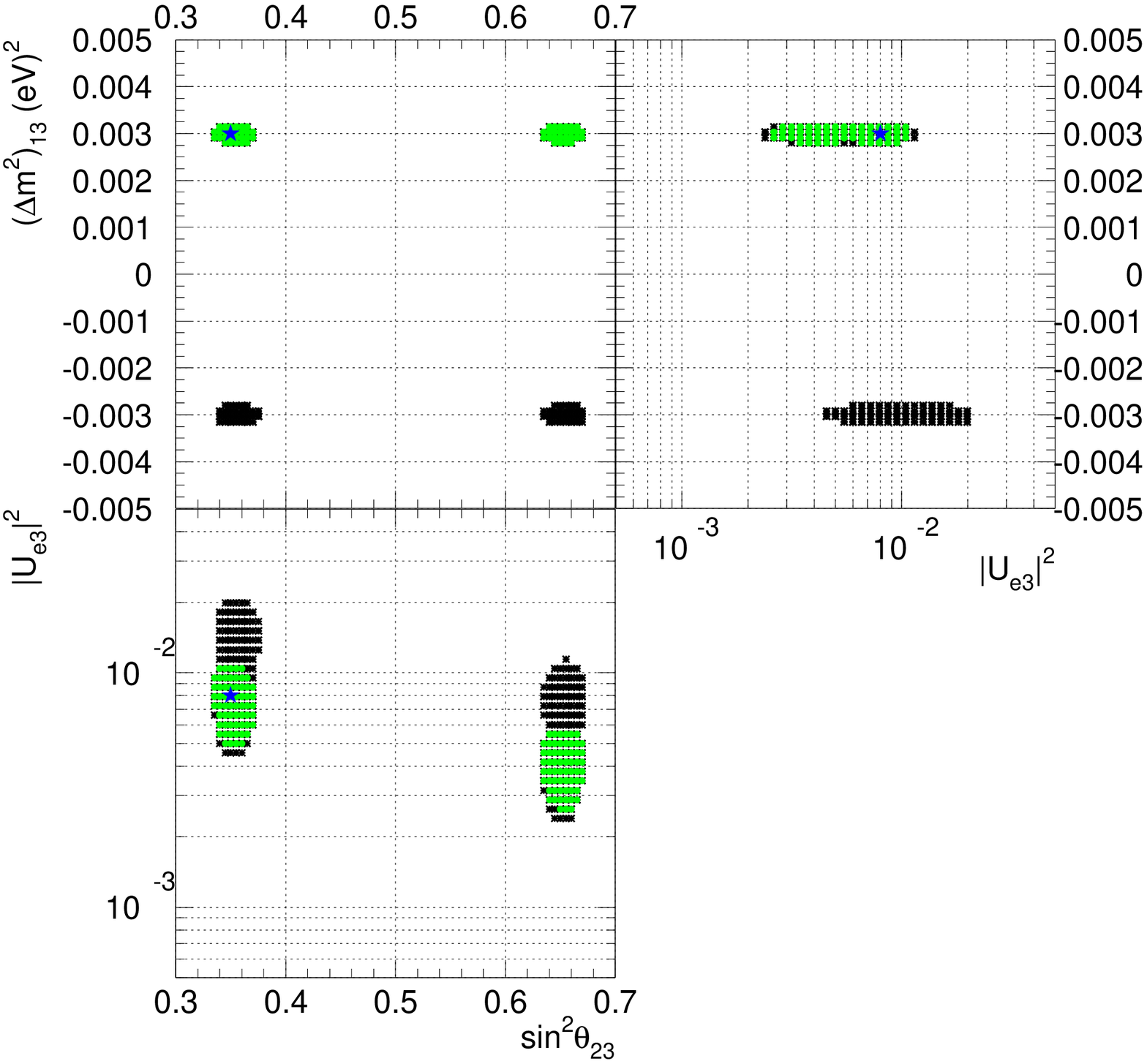}}
\caption{Same as Fig.~\ref{mea_05}, for $\sin^2\theta_{23}=0.35$.}
\label{mea_035}
\end{figure}

\begin{figure}[ht]
\vspace{1.0cm}
\centerline{\epsfxsize 14.2cm \epsffile{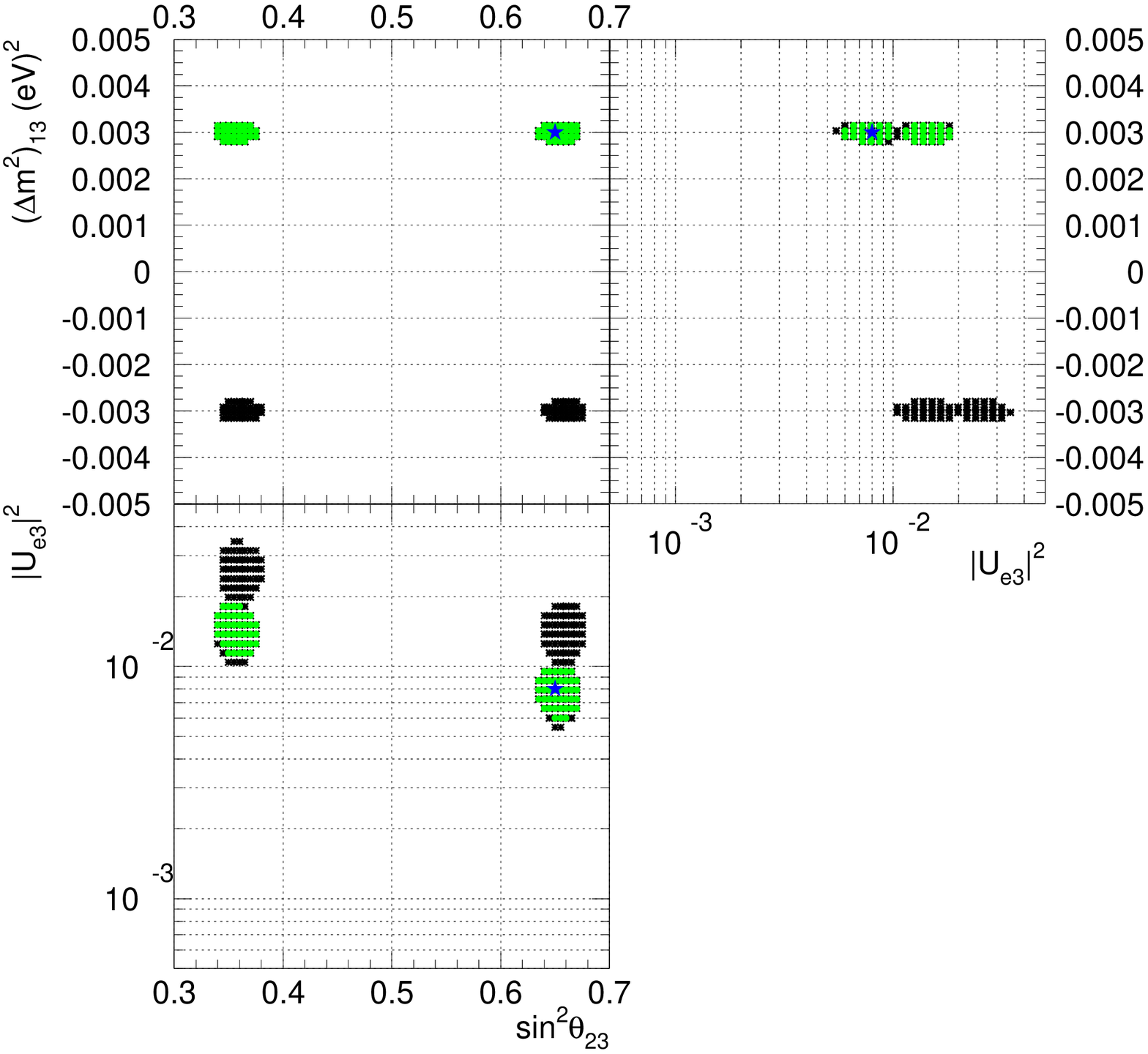}}
\caption{Same as Fig.~\ref{mea_05}, for $\sin^2\theta_{23}=0.65$.}
\label{mea_065}
\end{figure}

Once a signal is detected, the next step is to try and determine the neutrino mixing
parameters, $\Delta m^2_{13}$, $|U_{e3}|^2$ and $\sin^2\theta_{23}$. We do this for 
several different input values. The two-sigma confidence
level measurements in the  $(\Delta m^2_{13}\times|U_{e3}|^2)$, 
$(|U_{e3}|^2\times\sin^2\theta_{23})$, and $(\Delta m^2_{13}\times\sin^2\theta_{23})$-planes
obtained after 120 kton-years of ``neutrino-beam data'' are depicted in Figs.~\ref{mea_05},
\ref{mea_035}, and \ref{mea_065}, for $\sin^2\theta=0.5$, 0.35, and 0.65, respectively, 
$\Delta m^2_{13}=+3\times 10^{-3}$~eV$^2$ and $|U_{e3}|^2=0.008$. In the case of ``maximal''
atmospheric mixing (Fig.~\ref{mea_05}), there are still two disconnected solutions, 
centered around two distinct values of $\Delta m^2_{13}$ and $|U_{e3}|^2$. In this case,
one would obtain at the two-sigma level $0.005<|U_{e3}|^2<0.02$. For nonmaximal atmospheric
mixing (Figs.~\ref{mea_035}, \ref{mea_065}), the situation is even more confusing, and there
are four disconnected solutions centered at distinct values of all the three parameters of
interest.

Several other comments can be made regarding Figs.~\ref{mea_05}, \ref{mea_035}, and \ref{mea_065}.
First, as hinted in Sec.~\ref{sec_dis}, 
the size of the $\sin^2\theta_{23}$ uncertainty is significantly
larger for maximal atmospheric mixing than for the nonmaximal cases. Second, even though
$\sin^2\theta_{23}=0.35$ and $\sin^2\theta_{23}=0.65$ yield approximately 
the same value of $\sin^22\theta_{23}^{\rm eff}$ and therefore similar number of 
disappearing $\nu_{\mu}$s (see Sec.~\ref{sec_dis}), 
they yield very different number of appearing $\nu_e$s,
as discussed in Sec.~\ref{sec_app}. 
This fact is clearly visible if one compares Figs.~\ref{mea_035}, 
\ref{mea_065}. The error contours are smaller for $\sin^2\theta_{23}=0.65$ 
(Fig.~\ref{mea_065}) given the larger statistics. 
   
In order to ``solve'' the degeneracies, more information is needed. One option is to 
run with an antineutrino beam and keep the same experimental setup,
or perform a different experiment with a different baseline and neutrino beam. 
We will here pursue the first option, but warn readers that running with antineutrinos takes
a significantly larger amount of running time in order to obtain the same amount of data 
(especially if the neutrino mass hierarchy is normal). In our opinion, antineutrino
running is only realistic if an improved proton driver, such as the one currently being
investigated \cite{proton_driver} was present at NuMI. The existence of another experiment
with a different neutrino beam and baseline might turn out to be realistic if the JHF to
SuperKamiokande proposal materializes. 
    
We depict the two-sigma CL measurements in the  
$(\Delta m^2_{13}\times|U_{e3}|^2)$, $(|U_{e3}|^2\times\sin^2\theta_{23})$, and 
$(\Delta m^2_{13}\times\sin^2\theta_{23})$-planes obtained after 120 kton-years of 
``neutrino-beam data'' and 300 kton-years of ``antineutrino-beam data'' in Figs.~\ref{mea_05}, 
\ref{mea_035}, and \ref{mea_065}. One can
see that in the case of maximal atmospheric mixing (Fig.~\ref{mea_05}) the two-fold degeneracy
is lifted, and not only does one obtain a better measurement of $|U_{e3}|^2$ but, more
importantly (in our opinion), one can determine the neutrino mass-hierarchy. In the case
of non-maximal atmospheric mixing, the four-fold degeneracy collapses to a two-fold
degeneracy, and the neutrino mass hierarchy can also be unambiguously measured. 

\begin{figure}[ht]
\vspace{1.0cm}
\centerline{\epsfxsize 14.2cm \epsffile{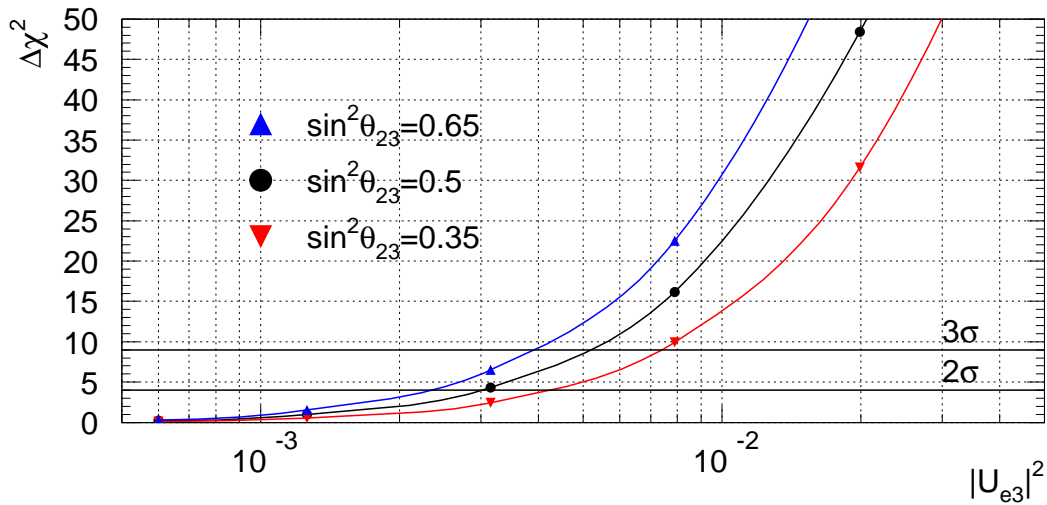}}
\caption{
$\Delta\chi^2\equiv|\chi^2_{\rm min}(\Delta m^2_{13}<0)-\chi^2_{\rm min}(\Delta m^2_{13}>0)|$
as a function of $|U_{e3}|^2$, for $\Delta m^2_{13}=+0.003$~eV$^2$ and $\sin^2\theta_{23}=
0.35,0.5,0.65$, obtained after 120~kton-years of ``neutrino beam data'' and 300~kton-years
of ``antineutrino beam data.'' See text for details. The solid horizontal lines indicate
the two- and three-sigma confidence levels for determining the sign of $\Delta m^2_{13}$.}
\label{hierarchy1}
\end{figure}

The capability of determining the neutrino mass hierarchy will depend on the size of 
$|U_{e3}|^2$. Fig.~\ref{hierarchy1} depicts the value of $\Delta\chi^2\equiv
|\chi^2_{\rm min}(\Delta m^2_{13}<0)-\chi^2_{\rm min}(\Delta m^2_{13}>0)|$
as a function of the input value of $|U_{e3}|^2$, assuming that the real value of 
$\Delta m^2_{13}=+3\times10^{-3}$~eV$^2$. $\Delta\chi^2$ is
defined after integrating out the marginal parameters $|U_{e3}|^2$ and $\sin^2\theta_{23}$,
and is to be interpreted as a $\Delta\chi^2$ for one degree of freedom 
($\Delta m^2_{13}$).\footnote{This is distinct from the analysis presented in \cite{our_paper}.} 
This means that, if $\sin^2\theta_{23}=0.5$, a two sigma hint that the neutrino mass
hierarchy is normal should be obtained if $|U_{e3}|^2=0.003$, while a five sigma discovery
of the neutrino mass ordering should be achieved if $|U_{e3}|^2\gtrsim 0.009$. The discriminatory
power depends on the value of $\sin^2\theta_{23}$, and increases with increasing
$\sin^2\theta_{23}$. As has been pointed out in Sec.~\ref{sec_app}, this is due to the fact that
the number of $\nu_e$ events due to oscillations is proportional to $\sin^2\theta_{23}$, and
more statistics can be obtained after the same running time for larger values of 
$\sin^2\theta_{23}$. Note that this is true even though the solution with $\sin^2\theta_{23}\neq 
0.5$ (statistically) is still ambiguous when it comes to determining the value of $|U_{e3}|^2$
and whether $\theta_{23}>\pi/4$ or $\theta_{23}<\pi/4$.

As mentioned in Sec.~\ref{sec_app}, solving the $\theta_{23}\leftrightarrow\pi/2-\theta_{23}$
degeneracy is much trickier. It is not resolved by comparing the neutrino channel
with the antineutrino channel, or even comparing different searches for $\nu_{\mu}
\rightarrow\nu_e$-transitions in long-baseline experiments, as long the matter effects
are far from the $\alpha\sim 1$ ``resonance'' region. In order to probe such a regime, however,
one is forced to probe very large neutrino energies, and hence very long neutrino baselines,
which implies the construction of new, futuristic, neutrino facilities. 

As we pointed out in Sec.~\ref{sec_app}, other options include looking for $\nu_{e}
\rightarrow\nu_{\tau}$-transitions \cite{silver}. In order to acomplish this, it is 
necessary to build
an intense, high energy source of electron-type neutrinos, {\it i.e.,}\/ a neutrino 
factory. One would have to decide, however, if the construction
of such a device is worthwhile given its physics capabilities in the scenario we are
considering here.\footnote{We do not advocate that there is no case for building a 
neutrino factory if the LMA solution is ruled out. We do believe, however, that 
its physics goals would have to be reviewed in detail. This statement, of course, 
only applies if no other new physics is discovered in the leptonic sector untill then.}

One final option would be to study the disappearance of electron-type (anti)neutrinos
from nuclear reactors. Unlike, say, the KamLAND experiment, such reactor would not
have to be a very long baseline setup, and CHOOZ-like distances
($O(1~{\rm km})$) would be optimal. The challenge would be, however, to improve the
``$\sin^22\theta$-sensitivity'' of the CHOOZ experiment by, say, a factor of 10,
from $O(0.1)$ to $O(0.01)$. Such an experiment would have to deal, for example, with
the issues which help define the sensitivity of the CHOOZ experiment, such as the overall
neutrino flux normalization, which is nominally known to a few percent \cite{chooz}. 
One way to try and address this issue would be to also include a near detector, {\it a la}\/ 
Bugey \cite{new_reactor}.    

\begin{figure}[ht]
\vspace{1.0cm}
\centerline{\epsfxsize 14.2cm \epsffile{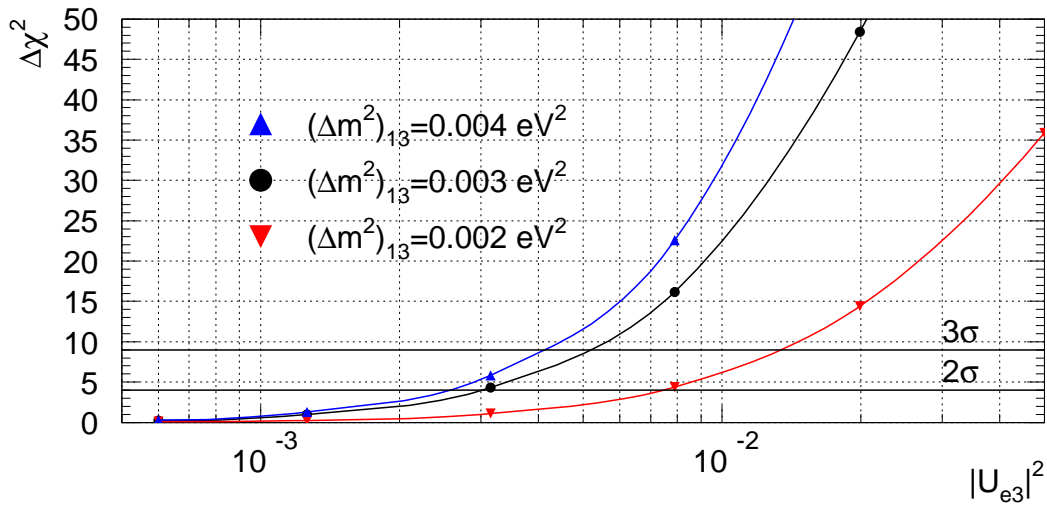}}
\caption{Same as Fig.~\ref{hierarchy1}, for $\sin^2\theta_{23}=0.5$ and 
$\Delta m^2_{13}=0.002,0.003,0.004$~eV$^2$.}
\label{hierarchy2}
\end{figure}

Next, we would like to address a more ``practical'' but important issue, related to how well
should the atmospheric parameters be measured before a ``safe'' decision regarding the 
location of the off-axis detector can be made. 
Part of the information required in order to address this issue is already contained in 
Fig.~\ref{sensitivity}. One sees that the sensitivity worsens significantly for
$|\Delta m^2|\lesssim 2\times10^{-3}$~eV$^2$ because the oscillation maximum begins to ``leave''
the energy window we are looking into, and the statistics begins to deteriorate significantly. 
A different question we would like to address is whether one can still try to obtain information
regarding the neutrino mass hierarchy. We do this by repeating the analysis depicted
in Fig.~\ref{hierarchy1} for fixed $\sin^2\theta_{23}=0.5$ but significantly 
different values of $\Delta m^2_{13}$. Fig.~\ref{hierarchy2} depicts the value of $\Delta\chi^2$
as a function of the input value of $|U_{e3}|^2$ for 
$\Delta m^2_{13}=+(2,3,4)\times 10^{-3}$~eV$^2$, which are meant to cover the currently 
allowed 99\% CL measurement of $|\Delta m^2_{13}|$. A more precise measurement should 
be expected after a few more years of K2K data \cite{k2k}, 
and will definitely be obtained after MINOS \cite{minos} and
the CNGS\cite{cngs} experiments ``turn on.'' 
 
One should note that, even for $\Delta m^2_{13}=0.002$~eV$^2$, 
where the statistics is expected to be significantly worse, a three sigma effect can be 
obtained for $|U_{e3}|^2\gtrsim 0.01$. For $\Delta m^2_{13}=0.004$~eV$^2$ it turns out that the
discriminatory power is larger than for $\Delta m^2_{13}=0.003$~eV$^2$, even though the
statistical sample is slightly smaller. This is due to the fact that for 
$\Delta m^2_{13}=0.004$~eV$^2$ the maximum of the oscillation takes place at a larger
value of the neutrino energy (compared with the ``optimal'' $\Delta m^2_{13}=0.003$~eV$^2$ ), 
such that when the mass hierarchy is inverted, the oscillation maximum starts to 
``leave'' the energy window we are looking into. This means that the number of events obtained
with the other choice neutrino mass-hierarchy is ``more different'' than in the case
$\Delta m^2_{13}=0.003$~eV$^2$, improving the discriminatory power. Incidently, the same 
phenomenon renders the discriminatory power in the case $\Delta m^2_{13}=0.002$~eV$^2$ worse.
Note that this effect came about because we chose the input value of $\Delta m^2_{13}$ to 
be positive. Were $\Delta m^2_{13}$ negative, we would have observed the opposite trend.

Finally, one may also inquire what happens if the baseline were different from the 900~km we
have chosen here. It turns out that for distances which are slightly smaller than 900~km
($L\gtrsim 700$~km) similar results can be obtained, as was discussed in detail in 
\cite{our_paper}. For significantly smaller baselines ($L\lesssim 500$~km), however,
the matter effects become less significant, and the ``optimal'' energy for a fixed value of 
$\Delta m^2_{13}$ becomes significantly smaller. This implies that while one might still
keep the sensitivity to observing $|U_{e3}|^2$ to similar levels, the power to 
discriminate between the two neutrino mass-hierarchies disappears. Furthermore, it 
should be added that performance for the iron-scintillator detector 
we are assuming for neutrino energies between 1 and 3~GeV does not apply for an equivalent
energy window centered below 1~GeV, as the capability to distinguish electrons from
neutral current events (and even muons) deteriorates. In this sub-GeV regime, one 
might be better off using different detector technologies, such as Water Cherenkov 
detectors (as is being proposed for the JHK to SuperKamiokande project)\footnote{We thank
Micha{\l} Szleper for comments concerning detector performance issues.}

\section{Concluding Remarks}

In the most conservative of scenarios, the fact that neutrinos have mass and mix will
be confirmed as the solution to the atmospheric and solar 
neutrino puzzles, while the LSND result 
will be ruled out as a signal for neutrino oscillations. In such a world, the main
goal of the next generation of neutrino oscillation experiments is very clear: measure
(or further constrain) the ``small'' $U_{e3}$ element of the leptonic mixing matrix.  
This is true irrespective of what the solution to the solar neutrino puzzle is. 
If the solar mass-squared difference is large enough, there is also the exciting
``bonus'' of trying to probe CP-violation in the leptonic sector. 

If the solution to the solar neutrino puzzle requires a very small $\Delta m^2_{\rm solar}$,
CP violation becomes inaccessible, as ``solar effects'' simply cannot be observed in 
terrestrial experiments. Nonetheless, we have shown that the physics prospects for
future long-baseline neutrino experiments are still very attractive. In many respects, 
measuring $U_{e3}$ becomes a simpler task, while several obstacles still remain. 
While overcoming these obstacles, we have shown that one can (must) not only measure
$|U_{e3}|^2$ in long-baseline experiments with non-negligible matter effects, but also 
determine the neutrino mass-hierarchy. Determining the mass hierarchy is arguably the second
clear goal of future neutrino experiments (whether or not neutrino oscillations are directly
involved) and just as exciting (in our opinion), as discovering CP-violation in the 
leptonic sector. Whether or not the neutrino mass-hierarchy is normal or inverted
is certainly cleaner to address if ``solar effects'' are not around to complicate things,
and requires information from two distinct oscillation channels, such as $\nu_{\mu}
\rightarrow\nu_e$-transitions at a long baseline, complemented by $\bar{\nu}_{\mu}
\rightarrow\bar{\nu}_e$-transitions at the same experiment. Another option is to 
combine a ``long'' and a ``short'' long-baseline search for $\nu_{\mu}
\rightarrow\nu_e$-transitions. This might well be the scenario that we will contemplate 
in the future if both the JHF to SuperKamiokande and the NuMI off-axis efforts
materialize.

We have included the effects of ``translating'' the information obtained in the 
disappearance channel to the appearance channel, error bars included. Not only can we
cleanly identify the confusion induced by not knowing what the neutrino mass-hierarchy is,
but also identify the $\theta_{23}$ versus $\pi/2-\theta_{23}$ degeneracy \cite{amb}, which proves
to be much harder to lift in the type of long-baseline experiments we (and most
of the community) imagine as feasible in the near to intermediate future. We do provide
a few candidate solutions, which involve new neutrino oscillation experiments, such as
a ``Super--Bugey'' effort \cite{new_reactor} 
or hunting for $\nu_e\rightarrow\nu_{\tau}$ transitions. Finally, it should be noted, of course,
that several of the issues we raise here are also applicable if the LMA solution is
confirmed by KamLAND.

In summary, the results of the KamLAND experiment (and, in a few years, of MiniBoone
\cite{miniboone})
will shape the future of neutrino physics experiments. If we are on the right track
({\it i.e.,}\/ if there are no major surprises or set-backs), 
however, all different roads seem to lead to a next-generation, 
long-baseline experiment capable of searching for $\nu_{\mu}\leftrightarrow\nu_e$-oscillations.
Even if the LMA solution is ruled out, several fundamental measurements can be performed
by such experiments, which are worthy of very serious consideration.  

\subsection*{Acknowledgments}
\noindent

This work was supported by the U.S.~Department of Energy
Grant DE-AC02-76CHO3000. It is a pleasure to thank Micha{\l} Szleper and
Mayda Velasco for collaboration in early stages of this work
as well as for ``experimental'' inputs throughout.
We are also grateful to Bob Bernstein for comments regarding the extraction
of the atmospheric parameters from the disappearance channel, to Maury Goodman for
carefully reading the manuscript and providing many important comments, and to
Hisakazu Minakata, for discussions regarding future reactor experiments and comments on the
manuscript.

\begin{appendix}

\section{Oscillation Probabilities and the LOW Solution}

The effective Hamiltonian that describes the 
time evolution of neutrinos 
in matter can be written in the flavor basis as 
\beq
H=\fr{1}{2 E}\left\{U
\left(\begin{array}{c c c} 0&&\\&\dm_{12}^2&\\&&\dm_{13}^2\end{array}\right)
U^\dagger  +
\left(\begin{array}{c c c} a&&\\&0&\\&&0\end{array}\right) \right\},
\eeq
where $a=G_F\sqrt{2}N_e 2E$ represents matter effects from the 
effective potential of electron-neutrinos with electrons, and 
$U$ is the flavor mixing matrix in vacuum (PDG representation),
\beq
U=\left (
\begin{array}{c c c}
c_{12} c_{13} & s_{12} c_{13} & s_{13} e^{-i\delta}\\
-s_{12}c_{23}-c_{12}s_{23}s_{13}e^{i\delta} & 
c_{12}c_{23}-s_{12}s_{23}s_{13}e^{i\delta} & s_{23}c_{13}\\
s_{12}s_{23}-c_{12}c_{23}s_{13}e^{i\delta} &
-c_{12}s_{23}-s_{12}c_{23}s_{13}e^{i\delta} & c_{23}c_{13}
\end{array}
\right ).
\eeq
By imposing $|\Delta m^2_{13}|>\Delta m^2_{12}>0$, we can relate $\theta_{23}$ and 
$|\Delta m^2_{13}|$ with the atmospheric angle and mass-squared difference, 
$\theta_{12}$ and $\Delta m^2_{12}$ with the solar angle and mass-square difference, and
$\theta_{13}$ with the ``reactor'' angle. These definitions are used throughout the paper 
(see \cite{our_paper} for details).

For a baseline $L$, the evolution of neutrino states is given by
\beq
\nu(L)=S(L) \nu(0),
\eeq
with
\beq
S(L)=e^{-i H L}
\eeq
for constant matter density. The corresponding effective Hamiltonian
for antineutrinos is obtained by $U \rightarrow U^*$ and
$a \rightarrow -a$.

If the Hamiltonian can be separated into $H=H_0+\epsilon H_1$,
where $H_0$ can be exactly solved and $\epsilon/ll 1$ such that
$H_1$ can be appropriately treated in perturbation 
theory, then $S(L)=S_0(L)+S_1(L)$ and
$S_0(L)=e^{-i H_0 L}$ gives the lowest order transition amplitudes.

For $|\dm_{12}^2| \;
\ll \; |dm_{13}^2|$, and  non negligible $a$, 
we choose 
\bea
H_0&=&\fr{1}{2 E}\left\{U
\left(\begin{array}{c c c} 0&&\\&0&\\&&\dm_{13}^2\end{array}\right)
U^{\dagger}+
\left(\begin{array}{c c c} a&&\\&0&\\&&0\end{array}\right) \right\} \nn\\
&=&\fr{1}{2E}\tilde{U}
\left(\begin{array}{c c c} 0&&\\&\dtm_2^2&\\&&\dtm_3^2\end{array}\right)
\tilde{U}^{\dagger},
\eea
where $\dtm^2$ describes the energy-level spacing in matter and
$\tilde{U}$ the effective mixing elements, and ignore $H_1$ effects, which are suppressed
by $\epsilon=\Delta m^2_{12}/\Delta m^2_{13}$. Of course,
\beq
S_0(L)=\tilde{U}
\left(\begin{array}{c c c} 0&&\\&e^{-i\fr{\dtm_2^2}{2 E}L}&\\
&&e^{-i\fr{\dtm_3^2}{2 E}L} \end{array}\right)
\tilde{U}^{\dagger}.
\eeq

The exact diagonalization of $H_0$ yields the effective mass differences 
and mixing angles in matter in the limit $|\dm^2_{12}/\Delta m^2_{13}|\ra0$, while 
no approximation have to be made regarding $a$. One obtains
\beq
\tilde{U}=\left(\begin{array}{ccc}
0&\fr{e^{-i\delta}}{n_2}(l_2-c_{13}^2)&\fr{e^{-i\delta}}{n_3}(l_3-c_{13}^2)\\
-c_{23}&\fr{1}{n_2}s_{23}s_{13}c_{13}&\fr{1}{n_3}s_{23}s_{13}c_{13}\\
s_{23}&\fr{1}{n_2}c_{23}s_{13}c_{13}&\fr{1}{n_3}c_{23}s_{13}c_{13}
\end{array}\right),
\eeq
for the mixing angles in terms of the vacuum parameters, where 
we have defined
\beq
n_2=\sqrt{l_2^2-2l_2c_{13}^2+c_{13}^2}; \quad
n_3=\sqrt{l_3^2-2l_3c_{13}^2+c_{13}^2}, 
\eeq
and 
\bea
l_2\equiv\fr{\dtm_2^2}{\dm_{13}^2}&=&
\fr{1}{2}\left[1+\alpha-
\sqrt{1+\alpha^2-
2\alpha\cos(2 \theta_{13}) }\right];\nn\\
l_3\equiv\fr{\dtm_3^2}{\dm_{13}^2}&=&
\fr{1}{2}\left[1+\alpha+
\sqrt{1+\alpha^2-
2\alpha\cos(2 \theta_{13}) }\right],
\eea
which give the effective mass differences.
We have introduced the dimensionless
parameter $\alpha\equiv\fr{a}{\dm_{13}^2}$
for the sake of simplicity. The ordering of levels in matter is
(1,2,3) for the hierarchical case of $\Delta m_{13}^2 > 0 $ and
$\alpha < 1$.

The effective mixing matrix $\tilde{U}$ 
is independent of $\theta_{12}$ and $\delta$, if 
the latter is appropriately rotated away.
The matrix $\tilde{U}$  can be expressed also 
in the PDG form.
To do so, we change
\beq
\tilde{U} \ra \tilde{U}\left( \begin{array}{ccc}
1&&\\&e^{i \delta}&\\&&1
\end{array}
\right ).
\eeq
By comparing with previous expressions, we may establish the
correspondence between the effective mixing angles, $\tilde{\theta}_{ij}$,
and the vacuum parameters, getting
\bea
\tilde{c}_{12}=0, \hspace{3cm} |\tilde{s}_{12}|=1, \nn \\
\tilde{c}_{23}=c_{23}, \hspace{3cm} \tilde{s}_{23}=s_{23}, \nn \\
\tilde{c}_{13}=\fr{l_2-c_{13}^2}{n_2}, \hspace{2cm}
\tilde{s}_{13}=\fr{l_3-c_{13}^2}{n_3},
\eea
up to signs. The vanishing mixing in matter $\tilde{c}_{12}=0$
is a consequence of the degeneracy $\Delta m^2_{12}=0$ in vacuum and
says that the lowest mass eigenstate in matter
contains no electron-neutrino flavor component.
This result is the ingredient that avoids genuine CP
violation in matter, even if one has three non-degenerate effective
masses.

The transition amplitudes for $\nu_\alpha \ra \nu_\beta$
in the $\dm^2_{12}/\dm^2{13}\rightarrow 0$ case are given 
by $S_0$ matrix elements, which can be written
\bea
A(\alpha\ra\beta;L)&=&S_0(L)_{\beta \alpha} \\
&=&
\delta_{\beta \alpha}+\tilde{U}_{\beta 2}\tilde{U}_{2\alpha}^\dagger
\left(e^{-i\fr{\dtm_2^2}{2 E}L}-1 \right)
+\tilde{U}_{\beta 3}\tilde{U}_{3\alpha}^\dagger
\left(e^{-i\fr{\dtm_3^2}{2 E}L}-1 \right). \nn
\eea
From this expression we may calculate all probabilities,
\bea
P\left(\nu_e \rightarrow \nu_e \right) &=& 1 - \sin^2(2\tilde{\theta}_{13})
 \sin^2\left[ \tilde{\Delta}_{13} \right] \\
P\left(\nu_\mu \rightarrow \nu_e \right) &=& s_{23}^2 
\sin^2(2\tilde{\theta}_{13})
 \sin^2\left[\tilde{\Delta}_{13} \right] \\
P\left(\nu_\tau \rightarrow \nu_e \right) &=& c_{23}^2 
\sin^2(2\tilde{\theta}_{13})
 \sin^2\left[\tilde{\Delta}_{13}\right] \\
P\left(\nu_\mu \rightarrow \nu_\mu \right)&=& 1 - s_{23}^4
\sin^2(2\tilde{\theta}_{13})\sin^2\left[\tilde{\Delta}_{13}\right] -
2 s_{23}^2 c_{23}^2 \left\{ 1 - \cos\left[\Delta_{13} (1 + \alpha) \right]
\cos\left[\tilde{\Delta}_{13}\right] 
\right. \nonumber \\
&+& \left. \cos(2\tilde{\theta}_{13})\sin\left[\Delta_{13} (1 + \alpha) \right]
\sin \left[\tilde{\Delta}_{13} \right] \right\} 
 \\
P\left(\nu_\tau \rightarrow \nu_\mu \right)&=&
s_{23}^2 c_{23}^2 \left\{ 2 - 2\cos\left[\Delta_{13} (1 + \alpha) \right]
\cos \left[\tilde{\Delta}_{13} \right] -
\right. 
 \sin^2(2\tilde{\theta}_{13})
 \sin^2\left[ \tilde{\Delta}_{13}  \right]  \nonumber \\
&+&\left. 2 \cos(2\tilde{\theta}_{13}) \sin\left[\Delta_{13} (1 + \alpha) \right]
\sin \left[ \tilde{\Delta}_{13} \right] \right\}
 \\
P\left(\nu_\tau \rightarrow \nu_\tau \right)&=& 1 - c_{23}^4
 \sin^2(2\tilde{\theta}_{13})
 \sin^2\left[ \tilde{\Delta}_{13} \right]  - 
2 s_{23}^2 c_{23}^2 \left\{ 1 - \cos\left[\Delta_{13} (1 + \alpha) \right]
\cos \left[\tilde{\Delta}_{13}\right] 
\right.  \nonumber \\
&+& \left. \cos(2\tilde{\theta}_{13})\sin\left[\Delta_{13} (1 + \alpha) \right]
\sin \left[\tilde{\Delta}_{13}\right] \right\} 
\eea
where 
\bea
\tilde{\Delta}_{13}\equiv  \Delta_{13}
\sqrt{1+\alpha^2-
2\alpha\cos(2 \theta_{13}) },
\; \; \; {\mbox{with}}\; \; \;
 \Delta_{13}\equiv \frac{\Delta m_{13}^2 L}{4 E} .
\eea 
and
\beq
\sin^2(2\tilde{\theta}_{13})=4\fr{s_{13}^2c_{13}^2}
{1+\alpha^2-
2\alpha\cos(2 \theta_{13})}.
\label{sin2t}
\eeq

In order to get the corresponding expressions for
antineutrinos, one simply exchanges $a \ra -a $, {\it i.e.},\/ $\alpha \ra -\alpha $. 
The effect of such a change in the probabilities comes from the
different relative sign between mass and matter terms in $H_0$.
It is important to note that, {\sl because we are disregarding $\Delta m^2_{12}$ effects}
(which is perfectly justified, as we are interested in the LOW solution)
the expressions for neutrinos and antineutrinos are exchanged when the neutrino mass
hierarchy is ``flipped'', {\it i.e.},
\begin{equation}
P(\nu_{\alpha}\ra\nu_{\beta})[\Delta m^2_{13}]
=P(\bar{\nu}_{\alpha}\ra\bar{\nu}_{\beta})[-\Delta m^2_{13}].
\end{equation}
This behavior is a consequence of the fact that we are, in practice, dealing with an
effective two-level system.

%

\end{appendix}


\begin{thebibliography}{10}

\bibitem{solar} Homestake Coll. (B.T.~Cleveland {\it et al.}),
Astrophys. J. {\bf 496}, 505 (1998);
GALLEX Coll. (W.~Hampel {\it et al.}\/),
Phys. Lett. {\bf B447}, 127 (1999);
V.N.~Gavrin, for the SAGE Coll.,
Nucl. Phys. {\bf B} (Proc. Suppl.) {\bf 91}, 36 (2001);
E.~Bellotti, for the GNO Coll.,
Nucl. Phys. {\bf B} (Proc. Suppl.) {\bf 91}, 44 (2001);
SuperKamiokande Coll. (S. Fukuda, {\it et al.}) Phys. Rev. Lett. {\bf
86}, 5651 (2001); {\bf 86} 5656 (2001); SNO Coll. (Q.R. Ahmad {\it et al}), 
Phys. Rev. Lett. {\bf 87}, 071301 (2001); {\bf 89}, 011301 (2002).  

\bibitem{SNO_solar} SNO Coll. (Q.R. Ahmad {\it et al}), Phys. Rev. Lett. {\bf 89}, 011302 (2002). 

\bibitem{atm} NUSEX Coll. (M. Aglietta {\it et al.}\/), Europhys.
Lett., {\bf 8}, 611 (1989);
R.~Becker-Szendy {\it et al.,}\/ Phys. Rev. {\bf D46},
3720 (1992);
Kamiokande Coll. (Y.~Hirata {\it et al.}\/), Phys.
Lett. {\bf B280}, 146 (1992); Kamiokande Coll. (Y.~Fukuda {\it et al.}\/),
Phys. Lett. {\bf B335}, 237 (1994);
Frejus Coll. (K.~Daum {\it et al.}\/), Z. Phys. {\bf C66}, 417 (1995);
W.A.~Mann, for the Soudan 2 Coll., Nucl. Phys. {\bf B}
(Proc. Suppl.) {\bf 91}, 134 (2001); B.C.~Barish, for the MACRO Coll.,
Nucl. Phys. {\bf B} (Proc. Suppl.) {\bf 91}, 141 (2001).

\bibitem{atm_sk} SuperKamiokande Coll. (Y. Fukuda {\it et al.}), Phys. Lett. {\bf B433}, 9 (1998);
Phys. Rev. Lett. {\bf 81},1562 (1998); {\bf 82}, 2644 (1999);
Phys. Lett. {\bf B467}, 185 (1999).  

\bibitem{lsnd} LSND Coll. (C.~Athanassopoulos {\it et al.}\/),
Phys. Rev. Lett. {\bf 75}, 2650 (1995); {\bf 77}, 3082 (1996); {\bf 81}, 1774
(1998); W.C.~Louis for the LSND Coll.,
Nucl. Phys. {\bf B} (Proc. Suppl.) {\bf 91}, 198 (2001);
LSND Coll. (A.~Aguilar {\it et al.}\/), hep-ex/0104049.

\bibitem{solar_fits} V.~Barger {\it et al.,}\/ Phys. Lett. {\bf B537}, 179 (2002);
P.~Creminelli, G.~Signorelli, and A.~Strumia, hep-ph/0102234 -- updated version (April 2002);
A.~Bandyopadhyay {\it et al.,}\/ hep-ph/0204286; 
J.N.~Bahcall, M.C.~Gonzalez-Garcia, and C.~Pe\~na-Garay, hep-ph/0204314;
P.~Aliani {\it et al.,}\/ hep-ph/0205053;
P.C. de Holanda and A.Yu.~Smirnov, hep-ph/0205241;
G.L.~Fogli {\it et al.,}\/ hep-ph/0206162;
M.B.~Smy, hep-ex/0108053.

\bibitem{general_fit} M.~Maltoni, {\it et al.}\/, hep-ph/0207227.

\bibitem{fit_strumia} A. Strumia {\it et al.,}
Phys. Lett.{\bf B541}, 327 (2002). See also G.F.~Fogli {\it et al.}\/ in \cite{solar_fits}.

\bibitem{Kamland} A.~Piepke for the KamLAND Collaboration, Nucl. Phys. Proc. Suppl. {\bf 91}, 
99-104 (2001);
S.~A.~Dazeley for the KamLAND Coll., arXiv:hep-ex/0205041.

\bibitem{Kam_sim} V.~D.~Barger, D.~Marfatia and B.~P.~Wood,
Phys.\ Lett.\ {\bf B498}, 53 (2001);
R.~Barbieri and A.~Strumia,
JHEP {\bf 0012}, 016 (2000).
H.~Murayama and A.~Pierce, Phys.\ Rev.\ {\bf D 65}, 013012 (2002).

\bibitem{Kam_sim_ex} A.~de Gouv\^ea and C.~Pe\~na-Garay, Phys.\ Rev.\ {\bf D 64}, 113011 (2001);
P.~Aliani {\it et al.,}\/ hep-ph/0207348.

\bibitem{nos} H.~Murayama and T.~Yanagida, Phys. Lett. {\bf B520}, 263 (2001);
G.~Barenboim {\it et al.,}\/, hep-ph/0108199;
G.~Barenboim, L.~Borissov and J.~Lykken, hep-ph/0201080;
A.~Strumia, Phys. Lett. {\bf B539}, 91 (2002).

\bibitem{Borexino} G.~Ranucci for the Borexino Coll., Nucl. Phys. {\bf B}
(Proc. Suppl.)\,{\bf 91}, 58 (2001).

\bibitem{borex_sim} A.~de Gouv\^ea, A.~Friedland and H.~Murayama,
Phys.\ Rev.\ {\bf D 60}, 093011 (1999); JHEP {\bf 0103}, 009 (2001).

\bibitem{lit1} V.~Barger, {\it et al.,}\/ Phys. Rev. {\bf D 22}, 2718 (1980);
A.~De Rujula, M.B.~Gavela, and P. Hernandez, Nucl. Phys.
{\bf B547}, 21 (1999);
P.~Lipari, Phys. Rev. {\bf D 61}, 113004 (2000);
S.~Dutta, R.~Gandhi, and B.~Mukhopadhyaya, Eur. Phys. J. {\bf C18}, 405 (2000);
D.~Dooling {\it et al.,}\/ Phys. Rev. {\bf D 61},073011 (2000);
M.~Freund {\it et al.,}\/ Nucl. Phys. {\bf B578}, 27
(2000);
I.~Mocioiu and R.~Shrock, Phys. Rev. {\bf D62}, 053017 (2000);
V.~Barger {\it et al.,}\/ Phys. Lett. {\bf B485}, 379 (2000);
Phys. Rev. {\bf D 62}, 073002 (2000);
{\bf D63}, 033002 (2001);
A.~Cervera {\it et al.,}\/ Nucl. Phys. {\bf B579}, 17 (2000), Erratum-ibid. {\bf B593},
731 (2001);
Z.-Z.~Xing, Phys. Lett. {\bf487}, 327 (2000); Phys. Rev. {\bf D63}, 073012
(2000);
A.~Bueno, M.~Campanelli, A.~Rubbia, Nucl. Phys. {\bf B589}, 577 (2000);
P.~Fishbane, Phys. Rev. {\bf D 62}, 093009 (2000); P.~Fishbane and P.~Kaus,
Phys. Lett. {\bf B506}, 275 (2001); P.~Fishbane and S.~Gasiorowicz,
hep-ph/0012230;
M.~Freund, P.~Huber, M.~Lindner, Nucl. Phys. {\bf B615}, 321 (2001);
J.~Arafune, J.~Sato, Phys. Rev. {\bf D 55}, 1653 (1997);
H.~Minakata, H.~Nunokawa, Phys. Rev. {\bf D 57}, 4403 (1998);
A.~Romanino, Nucl. Phys. {\bf B574} 675 (2000);
S.M.~Bilenky, C.~Giunti, W.~Grimus, Phys. Rev. {\bf D 58}, 033001 (1998);
K.~Dick {\it et al.,}\/ Nucl. Phys. {\bf B562}, 29
(1999);
M.~Tanimoto, Phys. Lett. {\bf B462}, 115 (1999);
A.~Donini {\it et al.,}\/Nucl. Phys. {\bf B574}, 23
(2000);
M.~Koike, J.~Sato, Phys. Rev. {\bf D 61}, 073012 (2000), Erratum-ibid.
{\bf D62}, 079903 (2000);
S.J.~Parke, T.J.~Weiler, Phys. Lett. {\bf B501}, 106 (2001);
T.~Miura {\it et al.,}\/ Phys. Rev. {\bf D 64}, 013002
(2001); M.~Koike, T.~Ota, J.~Sato, Phys. Rev. {\bf D 65}, 053015 (2002); 
P.~Lipari. Phys. Rev. {\bf D 64}, 033002 (2001);
J.~Burguet-Castell {\it et al.,}\/ hep-ph/0207080;
K~Whisnant, J.M.~Yang, and B.-L.~Yong, hep-ph/0208193;
M.~Aoki, K.~hagiwara, and N.~Okamura, hep-ph/0208223, 
and many more references therein.

\bibitem{our_paper} G. Barenboim {\it et al.,}\/ hep-ph/0204208.

\bibitem{JHF} {See http://neutrino.kek.jp/jhfnu, in
particular, Y. Itoh {\it et al.,} ``Letter of Intent: A Long Baseline Neutrino
Oscillation Experiment using the JHF 50 GeV Proton Synchrotron and the
Super-Kamiokande Detector'' (Feb. 2000); Y. Itoh {\it et al.,}\/ ``The JHF-Kamioka
Neutrino Project'', hep-ex/0106019.}

\bibitem{amb} V. Barger {\it et al.,}\/
Phys. Rev. {\bf D 63}, 113011 (2001); V. Barger, D. Marfatia, and K. Whisnant,
hep-ph/0206038; H.~Minakata, H.~Nunokawa, and S.~Parke, hep-ph/0208163.

\bibitem{chooz} M. Apollonio {\it et al.,}\/ Phys. Lett. {\bf B420}, 397 (1998);
Phys. Lett. {\bf B466}, 415 (1999).

\bibitem{investigation} See for example, several talks at the 
``NuFact'02 Workshop Neutrino Factories based on Muon Storage Rings,'' July 1--6, 2002,
Imperial College, London, http://www.hep.ph.ic.ac.uk/NuFact02/, and 
``New Initiatives for the NuMI beam at Fermilab,'' May 2--4, 2002, Fermilab, 
http://www-numi.fnal.gov/fnal\_minos/new\_initiatives/new\_initiatives.html.

\bibitem{k2k} S. H. Ahn {\it et al.,}\/ Phys. Lett. {\bf B511}, 178
(2001);  K.~Nishikawa, talk at the 
``XXth International Conference on Neutrino Physics and Astrophysics,'' May 24--30, 2002, Munich,
Germany, http://neutrino2002.ph.tum.de/ . See also G.~L.~Fogli, E.~Lisi and A.~Marrone,
Phys. Rev. {\bf D 65}, 073028 (2002).

\bibitem{minos}{See http://www-numi.fnal.gov/ .}

\bibitem{cngs}{See http://proj-cngs.web.cern.ch/proj-cngs/.}


\bibitem{proton_driver} G.~Barenboim {\it et al.,}\/ hep-ex/0206025.

\bibitem{silver} A.~Donini, D.~Meloni, and P.~Migliozzi, hep-ph/0206034.

\bibitem{new_reactor} K.~Inoue, H.~Minakata, F.~Suekane, H.~Sugiyama, and O.~Yasuda, 
to appear.  

\bibitem{miniboone}{See http://www-boone.fnal.gov/ .}

\end{thebibliography}
\end{document}